\documentclass[twocolumn,showpacs,superscriptaddress,preprintnumbers,amsmath,amssymb,pre]{revtex4}
\usepackage{graphicx}% Include figure files
\usepackage{dcolumn}% Align table columns on decimal point
\usepackage{bm}% bold math
\usepackage{amsmath}
\usepackage{epstopdf}
\usepackage{verbatim}

\begin{document}

\title{Anisotropy of Weakly Vibrated Granular Flows}

\author{Geert H. Wortel}
\affiliation{Huygens-Kamerlingh Onnes Lab, Universiteit Leiden, Postbus
9504, 2300 RA Leiden, The Netherlands}

\author{Martin van Hecke}
\affiliation{Huygens-Kamerlingh Onnes Lab, Universiteit Leiden, Postbus
9504, 2300 RA Leiden, The Netherlands}

\date{\today}

\begin{abstract}
We experimentally probe the anisotropy of the fabric of weakly vibrated, flowing granular media. Depending on the driving parameters --- flow rate and vibration strength --- this anisotropy varies significantly. We show how the anisotropy collapses when plotted as function of the driving stresses, uncovering a direct link between stresses and anisotropy. Moreover, our data suggests that for small anisotropies, the shear stresses vanish. Anisotropy of the fabric of granular media thus plays a crucial role in determining the rheology of granular flows.
\end{abstract}

\maketitle

For granular media, flow and spatiotemporal organization are intimately
connected. Flow induces dilation \cite{reynolds,losertosc,nedder},  mechanical agitations \cite{2012_prl_kamrin,kamrin, kamrin_secondary_rheology} and anisotropy \cite{losertosc,utter_rev,bob_aniso_nature,ew_aniso, hongkong,oda1,collins}, while in turn, packing density, agitations and anisotropy strongly effect granular rheology \cite{GDR,poulifric,pouliJRFM,kiri_prl,2012_pre_nichol,2012_prl_kamrin,kamrin, kamrin_secondary_rheology,pouliquen_couette,utter_rev,losertosc}.
Our understanding of this micro-macro coupling has made enormous steps, leading
to experimentally verified predictive theories for the steady flow of granular media \cite{GDR,poulifric,pouliJRFM,2012_prl_kamrin,kamrin, kamrin_secondary_rheology}.
For slow flows, agitations play a central role. Experiments
have shown that a primary flow in a granular medium completely changes the materials response, in particular erasing the yield stress
everywhere  \cite{kiri_prl,2012_pre_nichol,pouliquen_couette}. Combining these observations with insights for non-local flows of, e.g., emulsions \cite{bocquetnature,bocquetPRL}, a predictive nonlocal model for slow granular flows has appeared recently \cite{2012_prl_kamrin,kamrin, kamrin_secondary_rheology}.

What is missing is an understanding of non-stationary dense granular flows. In particular, shear reversal experiments indicate that
{\em anisotropy} has a profound effect: immediately after flow reversal, the shear stress drops to a low value and the system compacts, after which the system dilates and the shear stress increases back to its steady state value  \cite{losertosc,utter_rev}. Microscopically, symmetry-breaking force chains form during shear \cite{behringerandveje}, even when the real-space fabric remains
isotropic \cite{jaccoprl}. The qualitative picture that has emerged is that flow induces an anisotropic granular state which resist the flow, and then after reversal,
the medium needs to be sheared over a finite strain before it reorganizes into statistically steady state. The precise role of anisotropy and how strongly it influences granular flows remains unclear, and it is an open question how to experimentally probe the anisotropy in 3D flows.

Here we introduce an experimental protocol that allows us to both systematically
vary and probe the anisotropy in 3D granular flows. We employ our recently developed granular vibrheology system (Fig.~1(a)), which allows rheological measurements on weakly vibrated granular media in a split-bottom geometry \cite{splibo4,vibrheoprl,geertpre}.
The basic idea for probing the anisotropy is as follows: (I)
Prepare the granular medium by rotating the bottom disk, which is connected to a rheometer, at a constant rate. Externally supplied vibrations act as agitations, which tune the rheology \cite{vibrheoprl,geertpre} and, as we will show, the anisotropy as well. (II) Stop the flow, switch the rheometer from constant rate to constant torque mode, and then set the torque to zero. The driving disk then acts as a passive probe buried in a layer of anisotropic granular material.
(III) Vibrate the system, which leads to
counter-rotation of the disk due to the relaxation of the anisotropy of the medium.
We find that this rate can vary significantly, and use
it as a proxy for the anisotropy.

Using this protocol, we first find that variations
in the rheology go hand in hand with large variations of the anisotropy. Second,
for steady state flows, the relaxation of the anisotropy has a universal log-like shape. Third, we show that the data for the anisototropy collapses when plotted as function of the shear stress, revealing a deep link between anisotropy and resistance to flow. Finally, our data suggests that without anisotropy, grains would flow with very little resistance. Anisotropy is thus a crucial aspect of granular flows.

{\em Setup and Protocol:}
All our experiments are carried out in the same weakly vibrated split-bottom rheological cell as detailed in~\cite{geertpre} --- see Fig.~1a. Briefly, in this setup the rotation of a rough disk of radius $r_s=4$~cm at the bottom of a cylindrical cell drives a wide shear zone in the bulk of a layer (depth $H$) of a black soda-lime glass beads of diameter between 1 and 1.3~mm. This disk is coupled to a Anton Paar DSR 301 rheometer, allowing us to do experiments either at fixed torque $T$ or at fixed rotation rate $\Omega$.
In the absence of vibrations, the phenomenology of the flow is determined by the dimensionless filling height $h\equiv H/r_s$, which we fix at 0.33. In this regime, the shear bands are mainly vertical, and all grains above the disk co-rotate along with it (trumpet-flow)~\cite{splibo4}.
The experimental cell is vertically vibrated at a fixed frequency of 63~Hz by a VTS system VG100 shaker, where care is taken to decouple the rheometer from these vibrations \cite{geertpre}.
The amount of vibrations is characterized by the dimensionless parameter $\Gamma$=$A(2\pi f)^2/g$, where $A$ is the shaking amplitude and $g$ the gravitational acceleration  --- we focus on $0<\Gamma <1$.

\begin{figure}[t]
\begin{center}
\includegraphics[width=3.5cm,clip=true,trim=0cm 0cm 6.25cm 0cm]{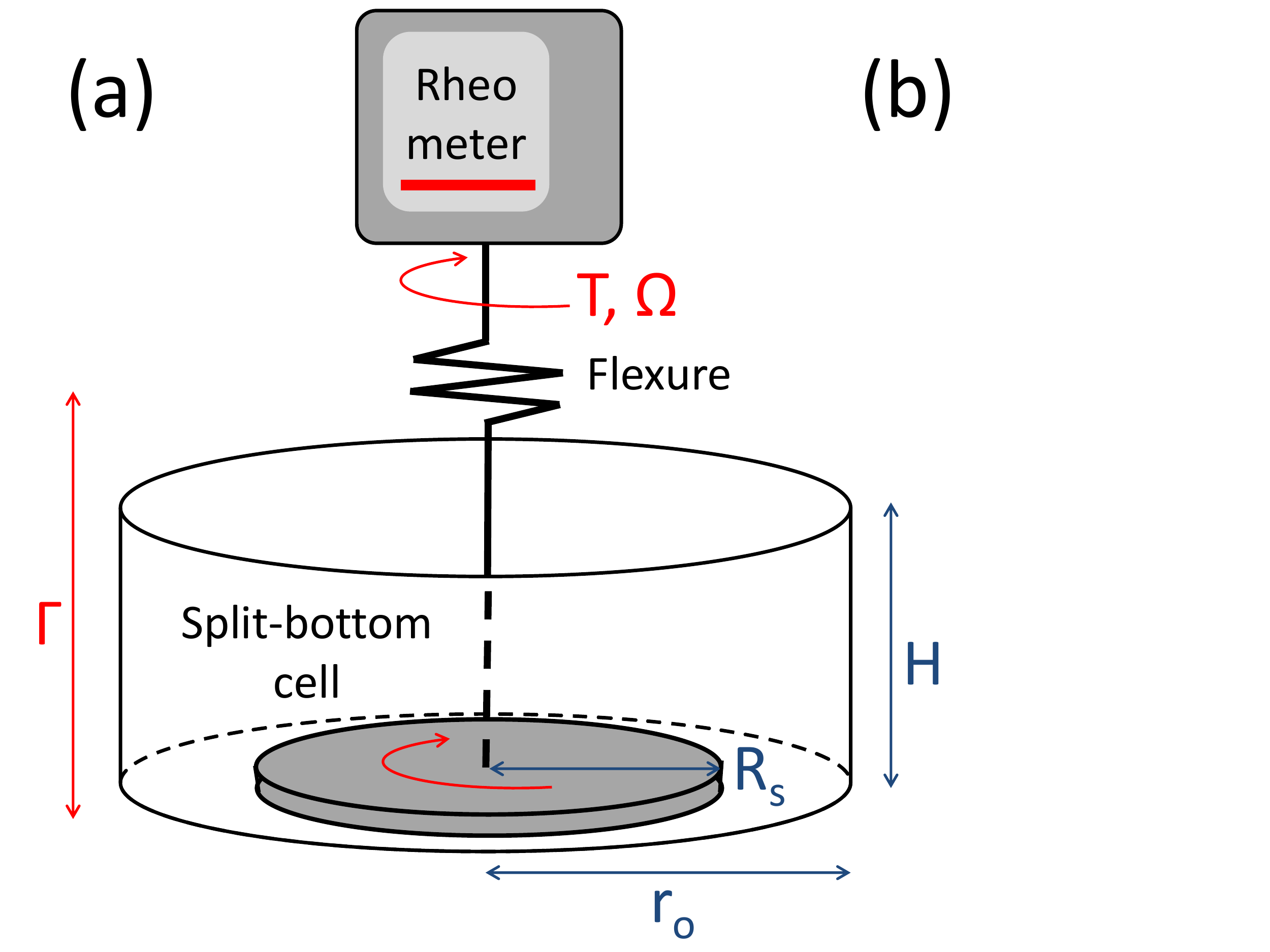}
\includegraphics[width=5cm]{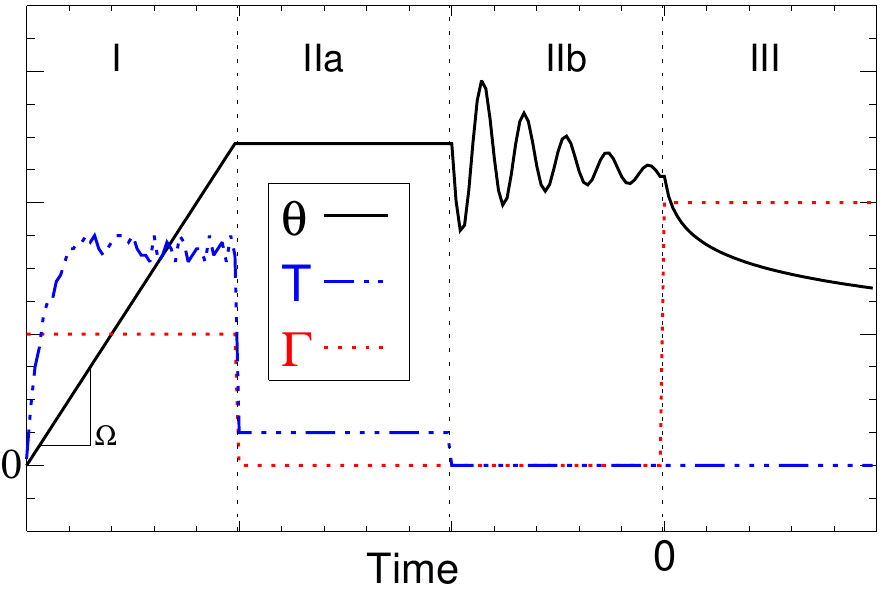}
\caption{{\footnotesize (color online) (a) Schematic picture of our vibrheological setup. (b) A schematic representation of the measuring protocol --- for details see text.}}\label{freeze4}
\end{center}
\end{figure}

The experimental protocol to measure the anisotropy is outlined in Fig.~\ref{freeze4}b, and consists of three main stages:
{\em(I) Preparation:} To start each experiment from a reproducible state, we start with preshearing the system (not shown in Fig.~1) by three consecutive
rapid (1~rps) shear motions in opposing directions, e.g., 5~s counter(co) clockwise, 10~s co(counter) clockwise and 5~s counter(co) clockwise, after which there is a
waiting period of 10~s during which the system is not sheared --- during the entire preshear stage, the vibrations are at $\Gamma$.
After this preshear, we prepare anisotropic states in
three series of experiments, that are aimed at
(a) determining at what strain the preshear becomes irrelevant, (b) testing whether
there are long-time memory effects, and (c) establishing the variation of the steady state anisotropy with $\Omega$ and $\Gamma$. Details of these protocols will be provided below.

{\em(II) Freeze:} To determine the anisotropy, we first need to stop the flow.
This part of the protocol is convoluted as care has to be taken to prevent disturbances of the anisotropy. For example,
abruptly changing the driving from finite $\Omega$ to $T=0$ would cause problems as the inertia of the disk would cause it to keep rotating, perturbing the packings anisotropy. Moreover, the flexure that couples disk and rheometer is under tension during stage {\em I} and this tension needs to relax without perturbing the packing. We therefore adopt a two step protocol: we first (IIa) switch from finite rate $\Omega$ to zero, and then immediately afterwards stop the shaking. The system is now frozen, with the anisotropy still present in the packing, but the flexure is tensed and the granulate is under stress.
We then (IIb) switch the rheometer to torque control and fix $T=0$. The flexure then experiences damped oscillations detected by the rheometer, and eventually relaxes. During these, the disk remains stuck in the sand, as all stresses encountered during the oscillation are less than the steady state stresses
in stage I, which, due to our use of finite $\Gamma$, are below the yield stresses for zero $\Gamma$. After these oscillations have damped out, stage II is finished, and the net shear stresses in the granular medium are zero.
We note that our measurements are robust with respect to changes in this part of the protocol (e.g., smoothly ramping down $\Gamma$ instead of a sudden stop, or stepping down $T$ in multiple steps).

{\em(III) Probe:}
To probe the anisotropy we now impose $\Gamma=\Gamma_{\mathrm{probe}}=0.4$, keeping
$T=0$. This leads to rotation of the disk, purely caused by the relaxation of the materials anisotropy. As schematically indicated in Fig.~1b, this relaxation is always opposite to the flow direction in stage I --- hence, a symmetry breaking field must be present to drive the systematic counterflow relaxation of the disk in stage III. We probe the angular coordinate of the rheometer $\theta(t)$ at a sample rate of 5~Hz for a total duration of 28~s, and note that since the rheometer's torque equals zero, the flexure remains relaxed and the rheometer's orientation and disk orientation remain equal, so that $\theta(t)$ describes the rotation of the buried disk. We note that other values of $\Gamma_{\mathrm{probe}}$ lead to similar conclusions --- the value of 0.4 gives reasonable relaxation times, and is thus well suited for our measurements.
We repeat all experiments five times and report averaged measurements, except for $\Omega < 10^{-5}$~rps where we average over three runs.

\begin{figure}[t]
\begin{center}
\includegraphics[width=\columnwidth]{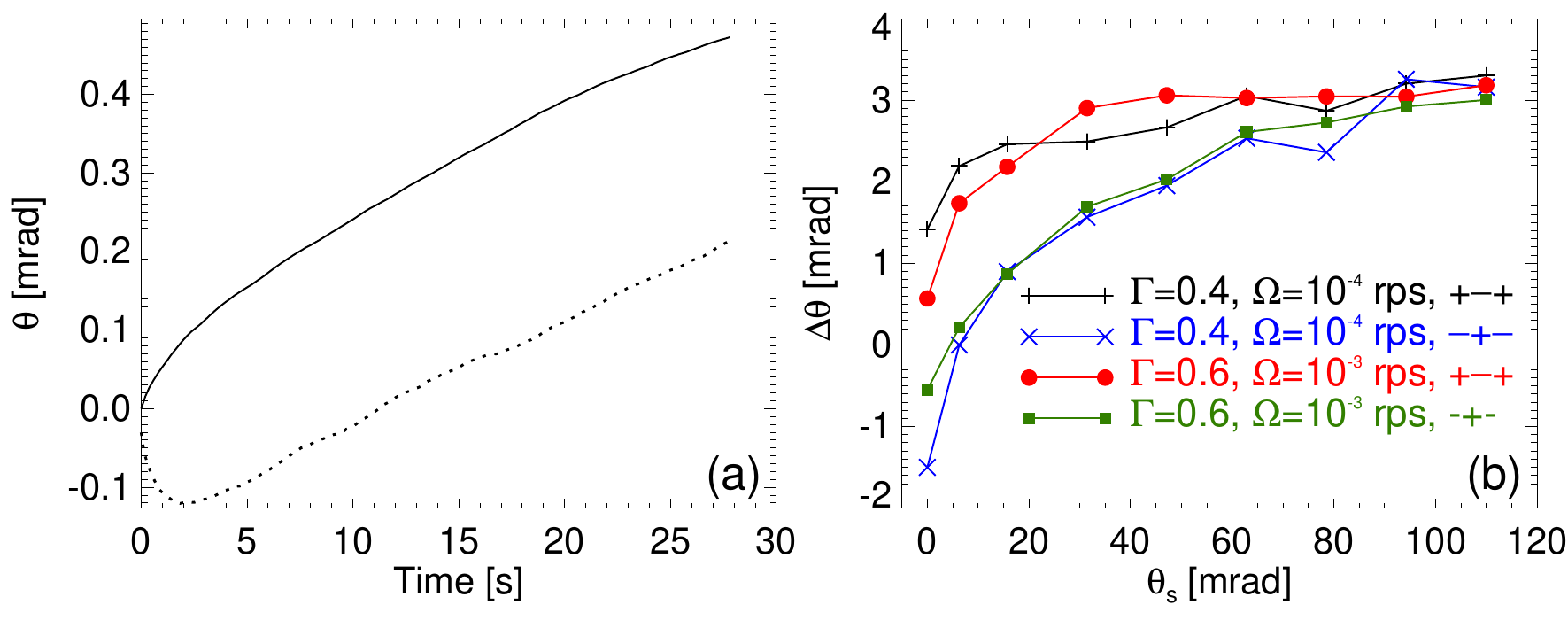}
\caption{{\footnotesize (color online) (a) Typical example of a relaxation curve for flow in steady state (solid line) for $\Gamma=1$, $\Omega=0.316\cdot10^{-4}$~rps and $\theta_s=100$~mrad. The dashed line reflects the non-monotonic transient case where the preshear was opposite to stage I for $\Gamma=0.6$, $\Omega=10^{-3}$~rps and $\theta_s=6.28$~mrad. (b) The dependence of $\Delta\theta$ on the (effective) strain $\theta:=t_s \times \Omega$ and the relative directions of preshear and evolution stage as indicated (where the rotation rate in the evolution stage is positive).
 }}\label{beatpreshear}
\end{center}
\end{figure}

{\em (a) Effect of preshear:} The packing that we create with the preshear protocol is well-defined and reproducible, but potentially strongly anisotropic.
To study the effect of preshear, we study $\theta(t)$ for two preshear protocols where the final preshear direction is either the same (+-+) or
opposite (-+-) to the positive shear direction used in the evolution stage I.

In Fig.~2a we compare $\theta(t)$ for opposite preshear directions.  Significant qualitative differences between these two cases can be seen. In the case of reversal, we observe  that $\theta(t)$ is non-monotonic --- this perhaps signifies that due to the inhomogeneous strain rates in the system, the relaxation also is spatially inhomogeneous.

We characterize such complex curves with $\Delta \theta := \theta(t=28s)$.
In Fig.~\ref{beatpreshear}(b) we show $\Delta \theta$  as function of the effective strain $\theta_s= 2 \pi t_s  \Omega$. Independent of rate and shaking strength, $\Delta \theta$ reaches a plateau for $\theta_{s}\gtrsim100$~mrad. Hence, for sufficiently large rotations, the anisotropy becomes independent of the preshear protocol.

{\em (b) Absence of memory effects:} We will now show that the steady state anisotropy is not
only independent of the preshear orientation, but that in general, the anisotropy goes to a well defined value. To do so, we have performed experiments where, after the preshear, the system is first driven at a fixed vibration intensity $\Gamma_1$ and rotation rate $\Omega_1$ for a fixed total rotation angle of the disk $\theta_1$, and then at $\Gamma_2$ and  $\Omega_2$ for a total disk rotation $\theta_2$.

\begin{figure}[t]
\begin{center}
\includegraphics[width=\columnwidth]{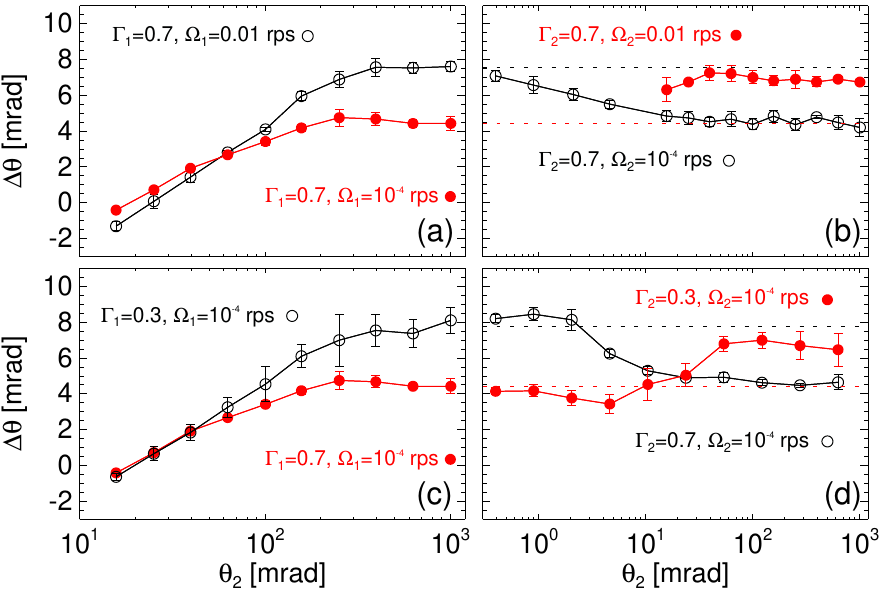}
\caption{{\footnotesize (color online)  The evolution of $\Delta\theta$ with $\theta_1$ and $\theta_2$. Going from the left to the right panels, the flow rate
$\Omega$ (a-b) or vibration strength $\Gamma$ (c-d) was changed. The relaxation of the data in panels (b) and (d) to the asymptotic values (dashed lines) obtained in panels (a) and (c) illustrates that for given ($\Omega,\Gamma$), $\Delta \theta$ relaxes to a unique value independent of the system's history. }}\label{doubledata}
\end{center}
\end{figure}

The results of such experiments are shown in Fig.~\ref{doubledata}.
For simplicity, $\Omega$ ($\Gamma$) is kept fixed, and
we interchange $\Gamma_1$ and $\Gamma_2$ ($\Omega_1$ and $\Omega_2$).
In Fig.~\ref{doubledata}(a-b) we show experiments for $\Gamma_1=\Gamma_2=0.7$, where $\Omega_1$ and $\Omega_2$ are $10^{-2}$ and $10^{-4}$ rps. In Fig.~\ref{doubledata}(a) we see how the anisotropy $\Delta \theta$ reaches a plateau, consistent with our earlier measurements. In Fig.~\ref{doubledata}(b) we see that if we interchange $\Omega_1$ and $\Omega_2$, the anisotropy evolves to essentially the same values as in Fig.~\ref{doubledata}(a): hence, after transients have died out, $\Delta \theta$ is only a function of $\Omega_2$. In Fig.~\ref{doubledata}(c-d) we show similar experiments for $\Omega$ fixed at $10^{-4}$~rps, and $\Gamma_1$ and $\Gamma_2$ at
0.3 and 0.7. Again, after transients have died out, $\Delta \theta$ is only a function of $\Gamma_2$. We have repeated such experiments also for a number of other parameter values, and always find the same behavior. Taken together, this shows that the anisotropy $\Delta \theta$ reaches a well defined steady state value, only
dependent on $\Gamma$ and $\Omega$.

{\em (c) Steady state anisotropy:}
We are now in a position to probe
the steady state anisotropy as a function of the main experimental parameters: the
flow rate $\Omega$ and the vibration intensity $\Gamma$. We vary
$\Gamma$ from 0.2 to 0.8 and have $\Omega$ span six decades, from
$10^{-6}$ to $1$ rps. We make sure the system is in steady state, which for the slowest $\Omega$ that we probe takes approximately 5 hours / per run.

\begin{figure}[t]
\begin{center}
\includegraphics[width=\columnwidth]{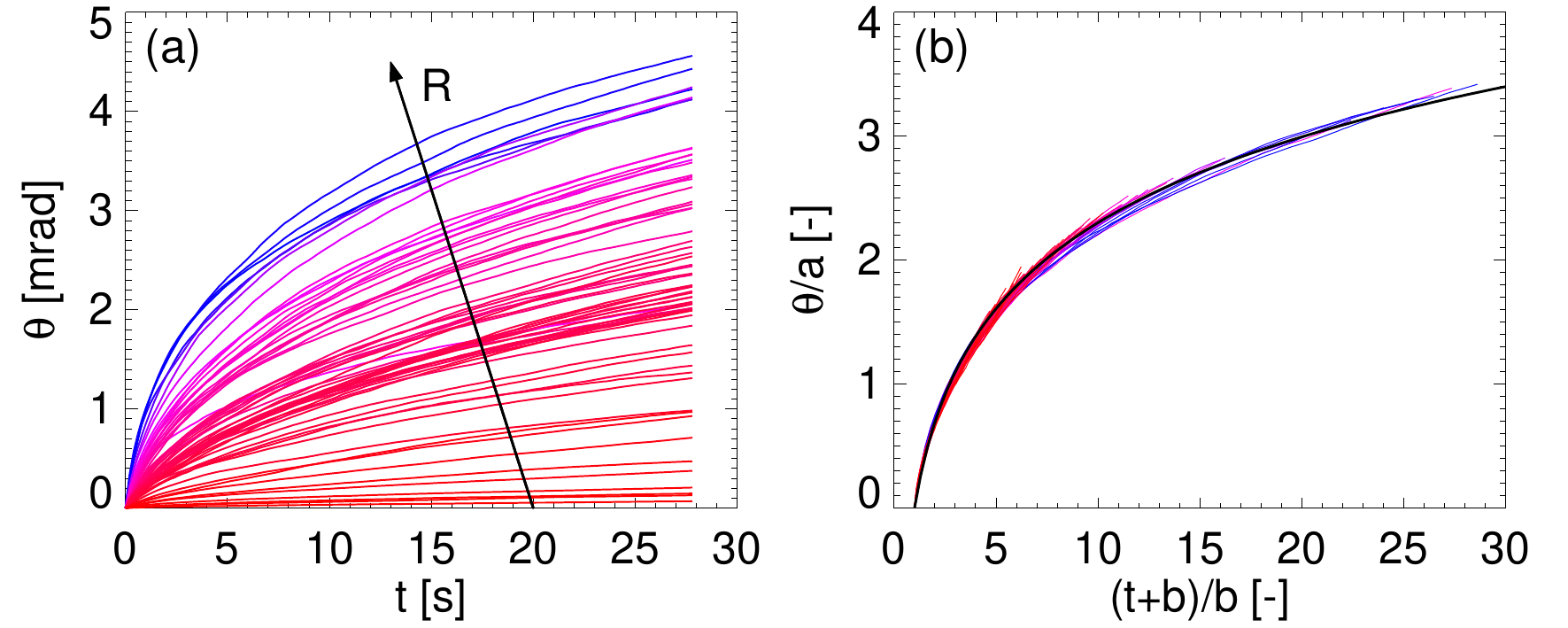}
\caption{{\footnotesize (color online) (a) A broad selection of $\theta(t)$ relaxation curves for the full range in $\Gamma$ and $\Omega$. From bottom (red) to top (blue), $R$ increases as the relaxation is faster. (b) A collapse of the data onto the master curve log[$(t+b)/b$] (plotted in thick black) using Eq.~\ref{relaxfit}.}}\label{relaxcollapse}
\end{center}
\end{figure}

In Fig.~\ref{relaxcollapse}(a) we present the relaxation curves $\theta(t)$ for all these experiments, where the color represents the relaxation speed at t=0~s. We observe that the magnitude of $\theta(t)$ varies over a large range, but that in this case where the system in stage (I) was in steady state, all curves have a similar shape reminiscent of a log.
As illustrated in Fig.~\ref{relaxcollapse}(b), we
find that this data can be fit very well by \cite{note}
\begin{equation}
\theta(t)=a ~\mathrm{log}\left(\frac{t+b}{b}\right)~. \label{relaxfit}
\end{equation}
As we are interested in the initial rate of change of $\theta$ as a proxy for the anisotropy in the frozen state, we define $R:=d \theta/dt|_{t=0}$, which using the fit to Eq.~\ref{relaxfit} can be accurately determined as $R=a/b$.

\begin{figure}[t]
\begin{center}
\includegraphics[width=\columnwidth]{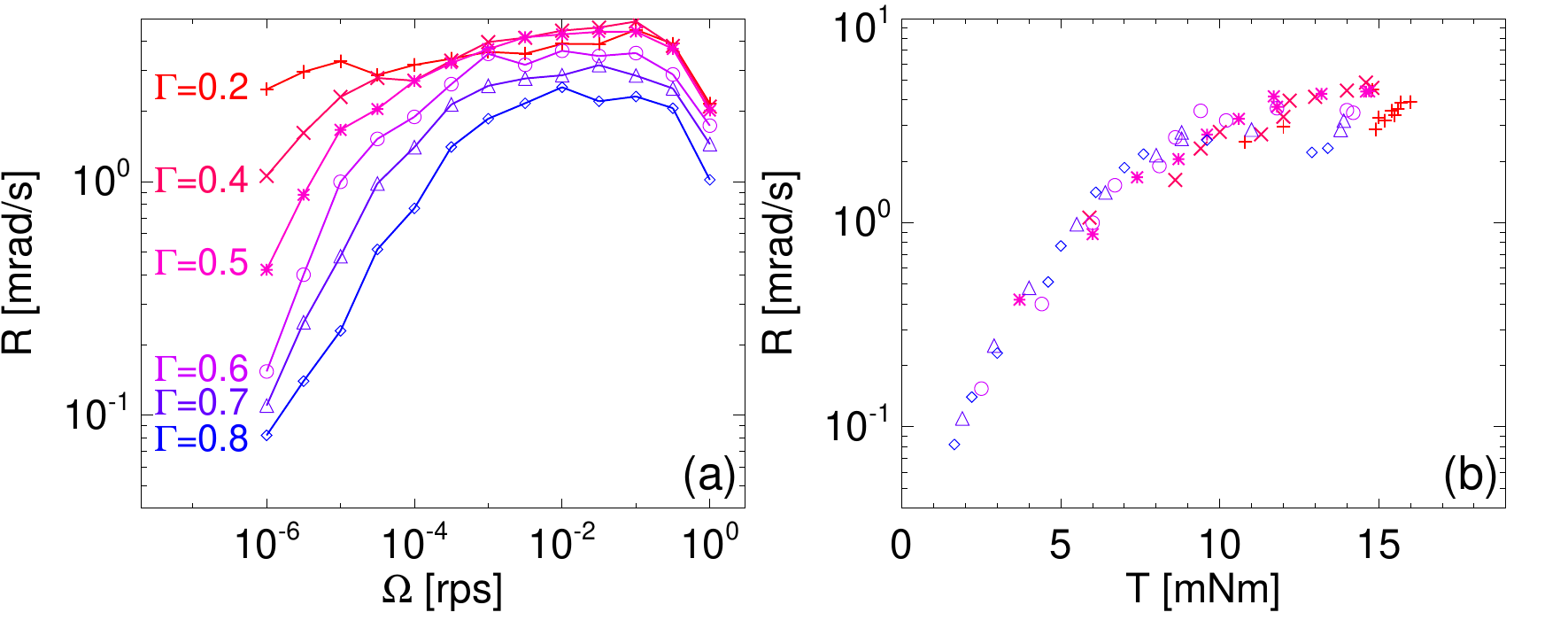}
\caption{{\footnotesize (color online) (a) The relaxation $R$ as a function of the control parameters $\Gamma$ and $\Omega$. (b) The data collapses when we plot $R$ as a function of the steady state driving torque $T$ measured during evolution stage 1.
}}\label{2paneltenn}
\end{center}
\end{figure}

In Fig.~\ref{2paneltenn}(a) we show $R$, showing significant variations with $\Omega$ and $\Gamma$.
To interpret these qualitatively, consider a competition between flow processes that generate anisotropy, and vibration induced relaxation of the anisotropy. In such a picture, the amount of anisotropy in the steady state will diminish with vibration magnitude, and this effect is strongest for slow flows where anisotropy generation is slow. In contrast, for fast flows, vibrations are not very effective in lowering the anisotropy and become essentially independent of vibration strength.

The variation of $R$ with $\Gamma$ and $\Omega$ is reminiscent to that of the driving torque $T$ \cite{vibrheoprl,geertpre}. For example, $R$ decreases monotonically with $\Gamma$ for low $\Omega$, varying over more than a decade --- in this same regime, the driving stresses also decrease monotonically with $\gamma$ \cite{vibrheoprl,geertpre}.  Moreover,
for low $\gamma$ and $10^{-4}<\Omega<10^{-1}$, $R$ is essentially independent of $\Omega$ --- precisely in this range, the flow is essentially rate independent \cite{vibrheoprl}. Finally, we note that  for $\Omega>0.1$~rps, $R$ decreases --- this is the transition to inertial flow where the stresses actually increase \cite{poulifric,vibrheoprl,geertpre}.

In Fig.~5(b) we plot $R$ as a function of $T$ (as measured in stage I), restricting ourselves to data for which $\Omega\leq 0.1$~rps. We observe a surprisingly good data collapse, without adjustable parameters. This shows that in the absence of inertial effects, anisotropy sets the granular rheology.

{\em Discussion and conclusion:} We will now briefly discuss the role of packing density. Notwithstanding the potential role of density, the qualitative trends in $R$ and $T$ are inconsistent with a purely density based picture. Specifically, suppose the anisotropy would be independent of the flow conditions in stage I (remember anisotropy in the granulate needs to be present to have the relaxation in stage III). If the variation of density would set the relaxation rate, a denser packing would relax slower than a looser packing. But this is inconsistent with  the behavior for low $\Omega$.
On the one hand, for stronger vibrations --- which lead to compaction of the material --- the relaxation indeed would be slower, consistent with our data. On the other hand, such reasoning would also suggest that for strong vibrations, the shear resistance $T$ goes up, but that is inconsistent with our data.

In contrast, a picture in which the variations in anisotropy are dominant makes more sense. Flow leads to the build up of anisotropy (Fig.~3a,c), whereas vibrations
lead to relaxation of the anisotropy --- counter to the previous flow direction, as observed here. Therefore, for given $\Omega$, stronger vibrations lead to more effective relaxation of the anisotropy, and a lower value of $R$ --- exactly as observed.
We thus conclude that anisotropy plays a dominant role in granular rheology.

In conclusion, we have introduced a novel experimental protocol to probe the anisotropy of granular media, by probing the relaxation of anisotropic packings under weak vibrations. Our data provides strong evidence that resistance to flow and anisotropy go hand in hand.  Moreover, our data suggests that in the absence of anisotropy, the resistance to flow is minute~\cite{losertosc}.
We note that the idea of a back stress,  generated in sheared granular materials, is reminiscent of kinematic hardening models~\cite{kine_har}, and suggest that such ideas may be a fruitful starting point for
more refined constitutive relations and models of non-stationary granular flows~\cite{poulifric,2012_prl_kamrin}. Finally, we suggest that anisotropy dominated effects are not limited to granular media, but may also be observable in other athermal, particulate media, such as  colloidal and macro-suspensions, foams and emulsions.

{\emph{Acknowledgments}---} We acknowledge discussion with J. Dijksman and O. Dauchot, and technical support from J. Mesman. GW acknowledges funding by FOM.
\bibliographystyle{apsrev}

\begin{thebibliography}{18}
\bibitem{note} The data must deviate from this fit at large $t$, but this is irrelevant for the determination of $R$.
\end{thebibliography}
{\footnotesize
{\hbadness 10000

}}

\end{document}